\documentclass{jps-cp}
\usepackage{txfonts} 
\usepackage{color}

\title{Floquet DMFT Analysis of High Harmonic Generation in Mott Insulators}

\author{Yuta \textsc{Murakami}$^{1}$, Martin \textsc{Eckstein}$^{2}$ and Philipp \textsc{Werner}$^{3}$}

\inst{$^{1}$Department of Physics, Tokyo Institute of Technology, Meguro, Tokyo 152-8551, Japan\\
$^{2}$Department of Physics, University of Erlangen-N\"urnberg, 91058 Erlangen, Germany\\
$^{3}$Department of Physics, University of Fribourg, Fribourg 1700, Switzerland}

\email{yuta.murakami@phys.titech.ac.jp}

\recdate{September 14, 2019}

\abst{We study the high harmonic generation (HHG) in Mott insulators using Floquet dynamical mean-field theory (DMFT). 
We show that the main origin of the HHG in Mott insulators is the doublon-holon recombination, and that the character of the HHG spectrum differs depending on the field strength. 
In the weaker-field regime, the HHG spectrum shows a single plateau as in the HHG from gases, and its cut-off energy $\epsilon_{\rm cut}$ scales linearly with the field strength $E_0$ as $\epsilon_{\rm cut}=\Delta_{\rm gap} + \alpha E_0$, where $\Delta_{\rm gap}$ is the Mott gap.
On the other hand, in the stronger-field regime, multiple plateaus emerge and the $m$-th cut-off scales as $\epsilon_{\rm cut,m}=U + m E_0$.
We show that this difference originates from the different dynamics of the doublons and holons in the weak- and strong-field regimes. 
We also comment on the similarities and differences between HHG from Mott insulators and from semiconductors.
This proceedings paper complements our recent work, Phys. Rev. Lett. {\bf 121}, 057405 (2018), with additional results and analyses.}

\kword{High Harmonic Generation, Strongly Correlated Systems, Nonequilibrium}

\begin{document}
\maketitle

\section{Introduction}
High harmonic generation (HHG) is a nonlinear phenomenon originating from strong light-matter coupling.
While HHG has been initially studied in gases and is technologically important\cite{Corkum1993PRL,Lewenstein1994}, 
recent observations of HHG in semiconductors open the possibility of HHG in condensed matter systems (CMs)\cite{Ghimire2011NatPhys,Schubert2014,Luu2015,Vampa2015Nature,Ndabashimiye2016,Liu2017,You2016,Yoshikawa2017Science}.
HHG signals from condensed matter can be more intensive than those from gases, which is useful for developing stronger lasers. The understanding of the HHG mechanism may also lead to spectroscopic techniques which yield information on the material and its electron dynamics.

Recently, HHG from CMs beyond semiconductors were observed in amorphous systems\cite{You2017Natcom,Luu2018Amorphas} and liquids \cite{Heissler2014NJP,Luu2018Liquid}.
These experiments demonstrate the possibility of HHG from correlated systems where the single-particle picture is not sufficient anymore and raise questions about the role of correlations in HHG
in CMs.
In order to understand the effects of correlations on HHG and explore new candidates for HHG sources, in this work, we theoretically study the HHG from Mott insulators (MIs), 
which is a prototypical class of insulating states in CMs.
By means of the single-band Hubbard model and the nonequilibrium extension of dynamical mean-field theory (DMFT), we reveal the characteristic features of the HHG in MIs, the main process for HHG, and the similarities and differences between MIs and semiconductors.

\section{Model and Method}

In this work, we focus on the Hubbard model at half-filling driven by an AC electric field,
\small
\begin{align}
H=&-\sum_{\langle i,j\rangle,\sigma} J_{ij}(t) c_{i,\sigma}^\dagger c_{j,\sigma}+U\sum_i n_{i\uparrow}n_{i\downarrow},
\end{align}
\normalsize
where $c^\dagger_{i,\sigma}$ is the creation operator of an electron at site $i$ with spin $\sigma$, $J_{ij}$ indicates the hopping parameter, and $U$ is the on-site interaction.
The effect of the external electric field is treated by the Peierls substitution with a vector potential ${\bf A}(t)$, $J_{ij}(t)=J_{ij}\exp\bigl(-iq\int^{{\bf r}_j}_{{\bf r}_i} d{\bf r} {\bf A}(t)\bigl)$.
Here $q$ is the electron charge.
We note that  ${\bf A}(t)$ is related to the electric field by ${\bf E}(t)=-\partial_t {\bf A}(t)$ and 
the setup is equivalent to a pure scalar potential term $\sum_{i,\sigma}\Phi({\bf r}_i,t) c^\dagger_{i\sigma}c_{i\sigma}=-q{\bf E}(t)\cdot (\sum_{i,\sigma}{\bf r}_i c^\dagger_{i\sigma}c_{i\sigma})$ in the Hamiltonian in a different gauge.
Furthermore, we take account of the effects of the environment by attaching a thermal bath $H_{\rm bath}$ to the Hubbard model.
Here we use a thermal bath of noninteracting electrons (the B\"{u}ttiker model) ~\cite{Tsuji2009,Murakami2017,Murakami2018PRL}.
Due to the coupling to a thermal bath, when the system is continuously excited by an external field with frequency $\Omega$,
it reaches a time-periodic nonequilibrium steady state (NESS) with a period $\mathcal{T}\equiv\frac{2\pi}{\Omega}$.

In the following, we calculate the NESS using the Floquet DMFT\cite{Aoki2013,Tsuji2009,Murakami2017,Murakami2018PRL,Murakami2018PRB}.
As in the usual DMFT, we map the original lattice problem to an effective impurity problem so that the impurity self-energy and the Green's function are identical to the local self-energy and the local Green's function of the lattice problem.
The Green's functions are formulated on the Keldysh contour since the NESS is determined by the balance between the excitation and the energy dissipation to the bath\cite{Aoki2013,Murakami2018PRB}.
To use DMFT, we focus on a hyper-cubic lattice with lattice spacing $a$ in the limit of infinite spatial dimensions ($J=\frac{J^*}{2\sqrt{d}}$ with $d\rightarrow\infty$), 
which has a Gaussian density of states $\rho(\epsilon)=\frac{1}{\sqrt{\pi}J^*}\exp[-\epsilon^2/J^{*2}]$.
We apply the field along the body diagonal, ${\bf A}(t)=A(t){\bf e}_0$ with ${\bf e}_0=(1,1,\cdots,1)$ and $A(t)=A_0\sin \Omega t$.
Hence the field strength along a given axis is $E(t)=-A_0\Omega \cos\Omega t\equiv -E_0\cos\Omega t$.
The B\"{u}ttiker model with a finite band width $W_{\rm bath}$ is used, 
$-{\rm Im}\Sigma^R_{\rm bath}(\omega)=\Gamma \sqrt{1-\left(\omega/W_{\rm bath}\right)^2}$, and 
 we set $q,a=1$ and use $J^*$ as the unit of energy. 
 
We consider parameters for which the Mott gap is larger than the Hubbard bands, 
typically $U=8,\beta=2.0,\Gamma=0.06,W_{\rm bath}=5,\Omega=0.5$.
Here $\beta$ is the inverse temperature of the bath.
This parameter choice justifies the use of the non-crossing approximatoin (NCA) and allows us to focus on the fundamental aspects of HHG in MIs.
The excitation frequency is chosen such that it is much smaller than the gap, as in the semiconductor experiments.
The HHG intensity is evaluated from the square of the Fourier transform of the dipole acceleration $\frac{d}{dt}j(t)$
as $I_{\rm hh}(n\Omega)=|n\Omega j(n\Omega)|^2$ with  $n\in \mathbb{N}$.
The current is $j(t)=iq \sum_{i,j,\sigma} v_{ij}(t)({\bf e}_0 \cdot {\bf r}_{i-j}) \langle c_{i,\sigma}^\dagger(t) c_{j,\sigma}(t)\rangle={\bf e}_0\cdot {\bf j}(t)$ and its Fourier component is $j(n\Omega)=\frac{1}{\mathcal{T}}\int^{\mathcal{T}}_0 d\bar{t} e^{i\bar{t}n\Omega} j(\bar{t})$.
Since we consider time-periodic NESSs, we only have 
a nonvanishing current 
at $n\Omega$ with $n\in\mathbb{N}$.

\section{Results}

\begin{figure}[tbh]
\includegraphics[width=140mm]{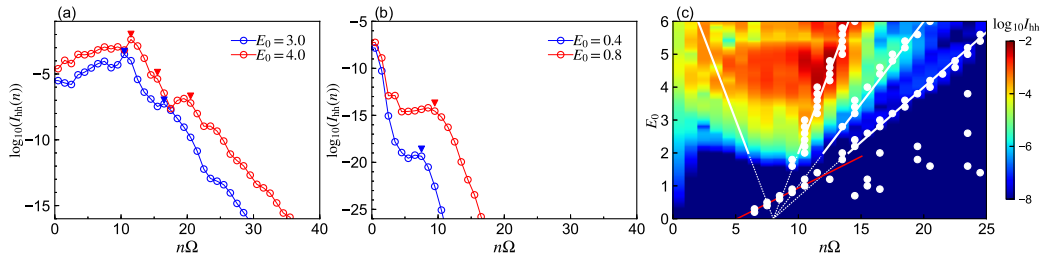}
\caption{(a,b) HHG spectra in the stronger-field regime (a) and in the weaker-field regime (b), and (c) HHG spectra as a function of the harmonic energy ($n\Omega$)
 and field strength ($E_0$). The arrows and the white circle markers show the cutoff energies. White lines in panel (c) indicate $n\Omega=U+mE_0$ and the red line is the fit for the weaker-field regime. We use $U=8.0,\beta=2.0,\Gamma=0.06,\Omega=0.5$.}
\label{fig:f1}
\end{figure}

In Fig.~\ref{fig:f1}, we show the behavior of HHG spectra from the MI  for various field strengths.
When the field is strong compared to the hopping ($J^*$), one can observe multiple plateaus, see Fig.~\ref{fig:f1}(a).
On the other hand, when it is comparable to or weaker than $J^*$, there is only one plateau as in atomic gases, see Fig.~\ref{fig:f1}(b).
The HHG spectrum is plotted in the plane of the harmonic frequency ($n\Omega$) and the field strength ($E_0$) in Fig.~\ref{fig:f1}(c).
As the field is increased, the cut-off energies of the plateaus in the HHG spectrum increase. 
Both in the weaker-field regime and the stronger-field regime, the cut-off energy ($\epsilon_{\rm cut}$) scales linearly with the field strength. More specifically, in the weaker-field regime $\epsilon_{\rm cut}\simeq \Delta + \alpha E_0$ with $\alpha>1$, while in the stronger-field regime we find 
$\epsilon_{\rm cut}\simeq U + m E_0$ with $m\in{\mathcal N}$.

\begin{figure}[tbh]
\includegraphics[width=140mm]{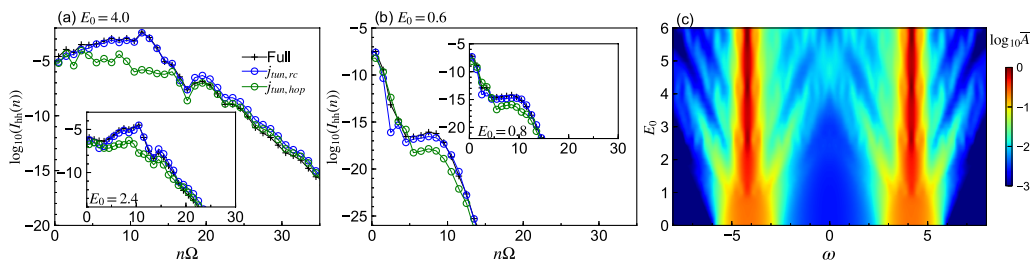}
\caption{(a)(b) Comparison of the HHG intensity extracted from the current originating from different processes in  the stronger-field (a) and  weaker-field (b) regimes.
  ``Full" corresponds to the full FDMFT calculation, $j_{\rm rc}$ is the current from the recombination processes, and $j_{\rm hop}$ is the current from the hopping processes of doublons and holons, 
  which are evaluated from the generalized tunneling formula presented in Ref.~\cite{Murakami2018PRB}. 
  (c) Time-averaged local spectral function $\bar{A}(\omega)$ of the nonequilibrium steady state as a function of $E_0$.}
\label{fig:f2}
\end{figure}

The first question we address is the main process responsible for the HHG in the MI.
In semiconductors, strong fields induce photo carriers (electrons in the conduction band and holes in the valence band).
The current in the semiconductors originates from two different processes, namely the intraband and the interband current \cite{Vampa2014PRL,Vampa2015PRB}.
The former corresponds to the dynamics of electrons and holes within the corresponding bands, while the latter is associated with the recombination of an electron and a hole.
In the MI, strong fields induce doubly occupied sites (doublons) and empty sites (holons), which act as photo carriers. 
As in the semiconductor case, the current can be separated into two contributions, the doublon/holon hopping process and the doublon-holon recombination process.
The former process does not alter the number of doublons and holons (the carrier number) and is analogous to the intraband current in semiconductors.
On the other hand, the latter process changes the number of doublons and holons and is analogous to the interband current in semiconductors.
Now the question is which process is dominant.
By means of the generalized tunneling formula derived in Ref.~\cite{Murakami2018PRB}, one can evaluate the currents from these two different processes.
The results are shown in Fig.~\ref{fig:f2}(a,b). 
In all regimes the recombination current dominates over the hopping current.
Therefore, we conclude that both in the weak- and strong-field regimes recombination is the dominant process for HHG.
We also note that, in some regimes, the contribution from the hopping process roughly follows that of the recombination process.
This is because, rigorously speaking, these processes cannot be fully decoupled since the states with different doulbon-holon numbers is slightly mixed via the finite transfer integral.  
(A similar effect has been reported in a semiconductor study under electric fields\cite{Vampa2014PRL}.)
In Fig.~\ref{fig:f2}(a) for $n\Omega>20$, we can see the values of these contributions become closer and cancel each other (destructive interference).

The next question is the origin of the different behavior depending on the field strength.
To find out the origin, we first show the (time-averaged) local spectrum of the NESS, $\bar A(\omega)\equiv -\frac{1}{\pi}{\rm Im} \bar{G}^R_{\rm loc}(\omega)$, see Fig.~\ref{fig:f2}(c).
Here $G^R_{\rm loc}$ indicates the retarded part of the local Green's function, and $\bar{G}^R_{\rm loc}(\omega)=\frac{1}{\mathcal{T}}\int^{\mathcal{T}}_0 dt_{\rm av}\int dt_{\rm r}G^R_{\rm loc}(t_{\rm r};t_{\rm av})e^{i\omega t_{\rm r}}$, where $t_{\rm r}$ denotes the relative time and $t_{\rm av}$ the average time.
One can see that for $E_0\lesssim J^*$, there is only a small renormalization of the Hubbard bands,
while for stronger $E_0$, they are strongly renormalized and one can observe Wannier-Stark side peaks.
Note that the width of the bands indicates the mobility of injected test charges.
Hence these features indicate that under the strong fields ($E_0\gtrsim J^*$) carriers tend to be localized.
These observations suggest the following picture.
In the stronger-field regime, after the doublons and holons have been created, they cannot move freely because of the field-induced localization.
At some point, they recombine by emitting light, which leads to the $U+mE(t)$ cutoff for the localized pairs of doublons and a holons separated by $m$ sites.
On the other hand, in the weaker-field regime, after the doublons and holons have been created, they can move 
more freely because of the not so strong field and acquire kinetic energy. 
In the recombination, they emit the minimum energy necessary to create a doublon and holon plus the kinetic energy.

\begin{figure}[t]
\includegraphics[width=140mm]{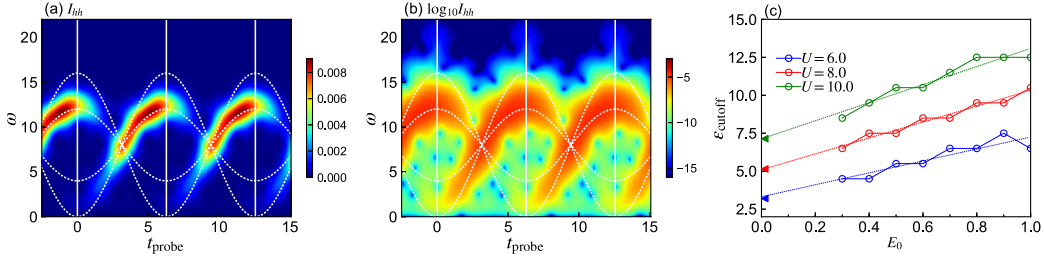}
\caption{(a,b) Normal-scale plot and log-scale plot of the temporal HHG intensity $I_{\rm hh}(\omega;t_{\rm probe})$ for $\Omega=0.5,E_0=4.0$. 
The dashed lines are $\omega=U\pm mE(t)$. Vertical lines indicate $t_{\rm probe}=0,\mathcal{T}/2,\mathcal{T}$.  
 Here $U=8.0,\beta=2.0,\Gamma=0.06$ and $W_{\rm bath}=5$.
 (c) Field-strength dependence of the cutoff energy in the weak-field regime for various $U$. Dashed lines are linear fits and the arrows at $E_0=0$ indicate the gaps estimated from the local spectral functions.}
\label{fig:f3}
\end{figure}

In the following, we provide additional data and analyses to support these scenarios. 
For the stronger-field regime, we perform a sub-cycle analysis to obtain a characteristic frequency 
emitted at each time ($t_{\rm probe}$) within one cycle \cite{Vampa2015PRB}.
More specifically, we calculate a windowed Fourier transformation of $j(t)$,  
$j(\omega;t_{\rm probe})=\int d\bar{t} e^{i\bar{t}\omega}j(\bar{t}) W(\bar{t};t_{\rm probe})$, 
and evaluate $I_{\rm hh}(\omega;t_{\rm probe})\equiv |\omega j(\omega;t_{\rm probe})|^2$.
Here $W(t;t_{\rm probe})$ is the Blackman window function with a half-window of length $2$ centered at $t=t_{\rm probe}$.
As is expected from the scenario of recombination of localized doublon-holon pairs, the signal follows $U\pm E(t)$ in the normal-scale plot Fig.~\ref{fig:f3}(a). This corresponds to the recombination of the nearest-neighbor pairs.
In the log-scale plot, Fig.~\ref{fig:f3}(b), the signal suddenly becomes weak for 
$|\omega|\gtrsim |U\pm 2E(t)|$, and in some time domain a clear strong signal is seen on the $|U\pm 2E(t)|$ curve.
 The latter correspond to the recombination of next-nearest-neighbor pairs.
These results support the scenario of localized pair recombination in the stronger-field regime.
We also note that this scenario can naturally explain the multiple plateaus and their scaling. The $m$-th neighbor pairs yield a signal at $U\pm mE(t)$, which oscillates from $U- mE_0$ to $U+ mE_0$.
Hence, the $m$-th plateau, whose edge scales as $U+ mE_0$, originates from the recombination of $m$-th neighbor pairs.

To analyze the weaker-field regime, we studied the cutoff energies for different values of $U$.
According to the scenario of the recombination of itinerant doublons and holons mentioned above, we expect that the emitted energy is the sum of the minimum energy necessary 
to create the doublon-holon pair (the Mott gap energy in the present case) and the kinetic energy which the carriers acquired under the influence of the periodic field ($E_{\rm kin}$).
In gas systems, the former corresponds to the ionization energy and the latter corresponds to the ponderomotive energy \cite{Corkum1993PRL,Lewenstein1994}.
Indeed, the offset $\Delta$, determined from extrapolations $E_0\rightarrow 0$, essentially coincides with the Mott gap ($\Delta_{\rm gap}$), which is determined from the local 
spectrum independently.  
Hence, it is consistent with the above scenario and our numerical results indicate that the acquired kinetic energy scales linearly against the field, $E_{\rm kin}=\alpha E_0$.

\begin{figure}[tbh]
\includegraphics[width=140mm]{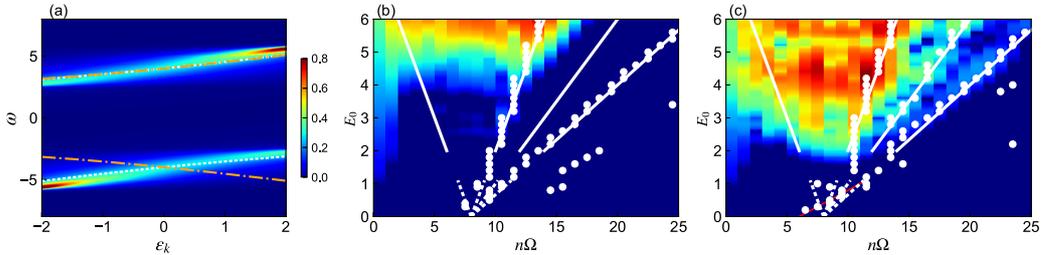}
\caption{(a) Momentum-dependent spectral function $A(\epsilon_{\bf k},\omega)$ of the Mott insulator in equilibrium. 
  The white lines show the peak position predicted by the H1A and the type 1 semiconductor, while the dashed orange lines show the dispersion of the type 2 semiconductor. (b)(c) HHG spectra $I_{hh}$ in the plane of $E_0$ and $n\Omega$. (b) is for the type 1 semiconductor model and (c) is for the type 2 semiconductor model. The color scale is the same as in Fig. 1 and $\Omega=0.5,U=8,\beta=2.0,\Gamma=0.06$.}
\label{fig:f4}
\end{figure}

Now we compare the HHG from MIs and semiconductors.
We consider a semiconductor model with a valence band and a conduction band, which corresponds to the 
upper and lower Hubbard bands, respectively:
\small
\begin{align}
H_{\rm semi}(t)
=-&\sum_{\langle i,j\rangle,\alpha}J^\alpha_{ij}(t) c_{i\alpha}^\dagger c_{j\alpha}
-\sum_{\langle i,j\rangle}J^{cv}_{ij}(t) (c^\dagger_{ic}c_{jv}+c^\dagger_{iv}c_{jc})+\sum_{i,\alpha}D_{\alpha}c^\dagger_{i\alpha}c_{i\alpha}.\label{eq:H_semicon}
\end{align}
\normalsize
$D_{\alpha}$ denotes the band center for band $\alpha=\{v, c\}$, and we choose $D_v=-U/2$ and $D_c=U/2$.
We introduce a transfer integral between the different semiconductor orbitals at neighboring sites, 
in order to mimic the fact that the hopping of electrons in MI leads to the creation of a doublon/holon pair on neighboring sites.
The effect of the electric field is included via the Peierls substitution
and we consider the NESS by attaching a B\"{u}ttiker-type thermal bath.
We consider two sets of hopping parameters, (i) $J^c=J^v=J^{cv}=0.5 J$ (we call this ``type 1" model) and (ii) $J^c=-J^v=J^{cv}=0.5 J$ (we call this ``type 2" model).
The type 1 model corresponds to the indirect gap semiconductors, whose dispersion is the same as that from the Hubbard 1 approximation (H1A), which is based on the atomic-limit self-energy $\Sigma^R(\omega)=\frac{U^2}{4\omega}$ \cite{Hubbard1963}.
Namely, for a momentum ${\bf k}$, the energy of the upper band and lower band is expressed as $\epsilon_{{\bf k},\pm}=(\epsilon_{\bf k}\pm \sqrt{\epsilon_{\bf k}^2+U^2})/2$ with the free particle energy $\epsilon_{\bf k} = -2J\sum_{a=1}^d \cos(k_a)$.
Indeed, the  dispersion of the single-particle spectrum obtained from DMFT follows nicely the dispersion of the type 1 model (or the H1A), see Fig.~\ref{fig:f4}(a).
On the other hand, the type 2 model corresponds to a direct gap semiconductor.
The HHG spectrum of the type 1 semiconductor is shown in Fig.~\ref{fig:f4}(b).
Intriguingly, the structure of the HHG spectrum turns out to be very similar to that from the H1A,
and one can observe cutoff energies that scale with $U+E_0$ and $U+3E_0$.
However, the model strongly underestimates the HHG spectrum in the weak to intermediate field regimes, 
because electron-hole pairs are not efficiently created. 
On the other hand, interestingly, the type 2 model shows qualitatively similar features (scaling and intensity distribution) as the MIs,
although its dispersion is very different from that of MIs, see Fig.~\ref{fig:f4}(a).
These comparisons demonstrate that the different nature of MIs and semiconductors is reflected in a very different relation between the HHG spectrum and the single-particle spectrum, and intuition from the semiconductor HHG does not work when one just looks at the single-particle dispersion.
We note that we have also studied the type1 semiconductor with disorders. The correlation from impurities produces a finite width in the spectrum at each ${\bf k}$ and makes the spectrum even closer to that of MIs, but such correlation effects do not result in a HHG spectrum which more closely resembles that of MIs.

\section{Conclusion}
In this proceedings, we analyzed the HHG from Mott insulating states described by the single band Hubbard model using the Floquet dynamical mean-field theory.
We showed that the main origin of the HHG is the doublon-holon recombination and that the character of the HHG spectrum changes due to the different dynamics of the doublons and holons in the weaker- and stronger-field regimes. We also discussed the similarities and differences between the HHG in Mott insulators and semiconductors.
Recent experiments have motivated numerous theoretical efforts to understand the HHG in correlated systems such as disordered systems \cite{Orlando2018,Ikeda2019}, strongly correlated systems \cite{Silva2018NatPhoton,Tancogne-Dejean2018,Ishihara2019} as well as spin systems \cite{Takayoshi2019PRB}.
We hope that our present work and these studies will stimulate further explorations of new HHG sources in consented matters.

\section*{Acknowledgements}
This work was supported by JSPS KAKENHI Grant Number JP19K23425.

\end{document}